# Clock Distribution and Readout Architecture for the ATLAS Tile Calorimeter at the HL-LHC

F. Carrió, *Member*, *IEEE*, and A. Valero on behalf of the ATLAS Tile Calorimeter System

*Abstract—* The Tile Calorimeter (TileCal) is one detector of the ATLAS experiment at the Large Hadron Collider (LHC). TileCal is a sampling calorimeter made of steel plates and plastic scintillators which are readout using approximately 10,000 PhotoMultipliers Tubes (PMTs). In 2024, the LHC will undergo a series of upgrades towards a High Luminosity LHC (HL-LHC) to deliver up to 7.5 times the current nominal instantaneous luminosity. The ATLAS Tile Phase II Upgrade will accommodate detector and Data AcQuisition (DAQ) system to the HL-LHC requirements. The detector electronics will be redesigned using a new clock distribution and readout architecture with a full-digital trigger system. After the Long Shutdown 3 (2024-2026), the on-detector electronics will transfer digitized hadron calorimeter data for every bunch crossing (~25 ns) to the Tile PreProcessors (TilePPr) in the counting rooms with a total data bandwidth of 40 Tbps. The TilePPrs will store the detector data in pipeline memories to cope with the new ATLAS DAQ architecture requirements, and will interface with the Front End LInk eXchange (FELIX) system and the first trigger level. The TilePPr boards will distribute the sampling clock to the on-detector electronics for synchronization with the LHC clock using high-speed links configured for fixed and deterministic latency. The upgraded readout and clock distribution strategy was fully validated in a Demonstrator system using prototypes of the upgraded electronics in several test beam campaigns between 2015 and 2018.

## I. INTRODUCTION

TILECAL [1] is the central hadronic calorimeter of the ATLAS experiment [2] at the Large Hadron Collider (LHC) at CERN. TileCal was designed to have a key role in the energy reconstruction of hadrons, jets, tau-particles and missing transverse momentum.

It is a sampling segmented calorimeter using steel plates as passive absorber and plastic scintillator tiles as active material. The TileCal detector is composed of four cylinders subdivided azimuthally into 64 wedge modules each. The two central cylinders form the Long Barrel (LB) while the other two cylinders, located on each side of the LB, are called the Extended Barrels (EB) (Figure 1). The modules in the LB are equipped with up to 45 PMTs whereas 32 PMTs are needed to read out the EB modules. The on-detector electronics are installed in mechanical structures, called super-drawers, housed inside girders in the outermost part of the modules.

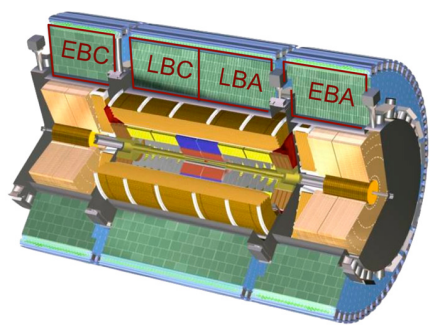

Fig. 1. Tile Calorimeter detector.

The light produced by the particles crossing the detector is read out using up to 9852 PhotoMultiplier Tubes (PMTs) for the entire detector. The signals from the PMTs are amplified, shaped, and digitized at 40 Msps using a clock synchronous with the LHC beam crossing (Figure 2).

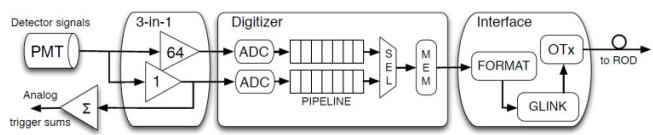

Fig. 2. Block diagram of the current readout architecture.

The digital samples are stored in pipeline memories during the Level-1 trigger latency (2.5 µs). Simultaneously, the PMT analog signals are transmitted to the Level-1 Calorimeter trigger system. The digital samples corresponding to the events selected by the Level-1 trigger system are transmitted through optical fibers to the ReadOut Drivers (RODs) [3] located off-detector at a maximum trigger rate of 100 kHz in average.

The custom 9U VME ROD module is the core element of the off-detector electronics of the current system. Each ROD board can read out up to 8 detector modules. Thus, 32 ROD modules installed in four VME crates, corresponding to the four detector barrels, are used to operate the entire detector. The main tasks of the RODs, executed in Digital Signal Processors (DSPs), are energy and time reconstruction, synchronization of detector and trigger data, busy handling and data compression.

The ROD receives 7 digital samples for each event selected by the Level-1 trigger system, and processes them within 10 µs to minimize the dead-time in the detector. Then, the processed data are transmitted to the Level-2 trigger system located in the High Level Trigger (HLT).

## II. ATLAS Phase II Upgrade

The LHC has planned a series of upgrades culminating in the High-Luminosity LHC (HL-LHC) which will increase the instantaneous luminosity up to $7.5 \cdot 10^{34}$ cm$^{-2}$s$^{-1}$ with a pileup close to 200 collisions per bunch interaction providing a total integrated luminosity of 4000 fb$^{-1}$ in ten years. In order to cope with the new radiation levels and increased data bandwidths for the HL-LHC, the TileCal readout electronics will be redesigned introducing a new readout strategy with a full-digital trigger system.

After the ATLAS Phase II Upgrade [4], the TileCal on-detector readout electronics will transmit digital detector data for every bunch crossing (~25 ns) to the Tile PreProcessor (TilePPr) boards in the counting rooms. As described in Table I, this new readout strategy implies an important increase of the required bandwidth for the communication between on- and off-detector electronics and trigger systems, keeping the same ratio between the number of readout channels and off-detector electronics boards.

TABLE I
COMPARISON OF THE TILE CALORIMETER READOUT PARAMETERS

|  | Current | HL-LHC |
|---|---|---|
| Total BW | 205 Gbps | 40 Tbps |
| Nb. fibers | 256 | 4096 |
| BW/module | 800 Mbps | 160 Gbps |
| Nb. boards | 32 | 32 |
| Nb. crates | 4 (VME) | 4 (ATCA) |
| In BW/board | 6.4 Gbps | 1.28 Tbps |
| Out BW/board$_{DAQ}$ | 3.2 Gbps | 40 Gbps |
| Out BW/board$_{Trigger}$ | Analog | 500 Gbps |

All detector data received from the on-detector electronics will be stored in pipeline memories until the reception of a trigger acceptance signal. Selected data will be transmitted from the TilePPr board to the ATLAS Data AcQuisition (DAQ) system through the Front End LInk eXchange (FELIX) [5]. While storing the data in the pipelines, the TilePPr must provide reconstructed cell energy for every bunch crossing to the ATLAS trigger system with a fixed and maximum latency of 1.7 µs [4].

In addition, in the new readout architecture the TilePPr board will distribute the LHC clock to the on-detector electronics for the digitization of the PMT signals and for the synchronization with the rest of the readout electronics of the DAQ system. Figure 3 shows a sketch of the readout architecture envisaged for the TileCal detector at the HL-LHC.

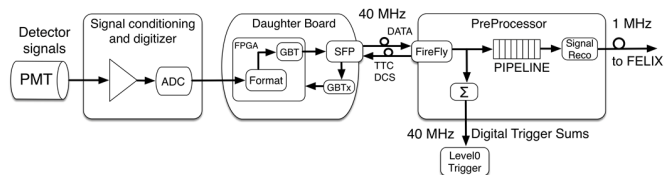

Fig. 3. Block diagram of the TileCal readout architecture at the HL-LHC.

## III. Demonstrator Upgrade Project

As a part of the evaluation of the readout architecture envisaged for TileCal in the HL-LHC, the TileCal collaboration built a fully functional module equipped with the upgraded electronics, called the Demonstrator module. The Demonstrator module was evaluated during different test beam campaigns between 2015 and 2018 in the H8 beam line of the SPS accelerator at CERN.

The test beam setup is combined with two more TileCal modules: one EB module and one LB module. These modules were instrumented with the legacy readout electronics in order to compare the performance of the upgraded and legacy electronics. Figure 4 shows a picture of the complete test beam setup where the three TileCal modules are piled up on a movable table capable of placing modules at any combination of angle and position with respect to the incident beam.

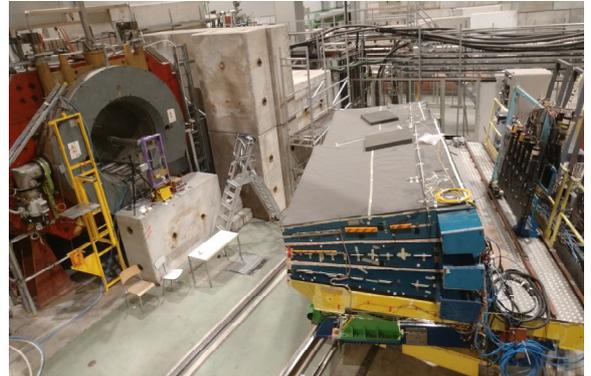

Fig. 4. Picture of the test beam setup at the SPS accelerator facilities.

The Demonstrator project plans include the replacement of one of the TileCal modules in ATLAS by the Demonstrator module during the Long Shutdown 2 (2019-2021). For this reason, the Demonstrator module was designed with the capability of providing both analog trigger signals to the current ATLAS Level-1 trigger system and full-digital trigger data for the upgraded system.

### A. Upgraded on-detector electronics

The Demonstrator module includes all the upgraded on-detector electronics required for the acquisition of the PMT signals, high-speed interface with the off-detector electronics and the high voltage distribution system for the PMTs.

The TileCal modules for the HL-LHC will be segmented into four identical and independent modules, called mini-drawers, to improve the reliability of the system through redundancy. Figure 5 shows a drawing of a mini-drawer with the position of all the on-detector components. Each mini-drawer is composed of a mechanical aluminum structure, one Main Board [6], one Daughter Board [7] and one HV regulation board to read out and operate up to 12 PMT blocks equipped with upgraded 3-in-1 Front End Boards (FEB) [6].

Each Main Board digitizes the analog signals received from up to 12 PMT blocks. The Main Board hosts a Daughter Board, which collects and transmits digitized data to the TilePPr prototype in the off-detector electronics at the LHC frequency.

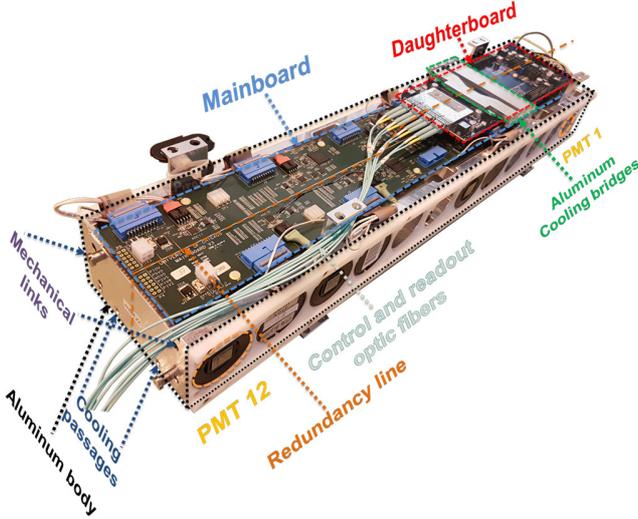

Fig. 5. Drawing of the mini-drawer with the position of the different elements of the on-detector electronics.

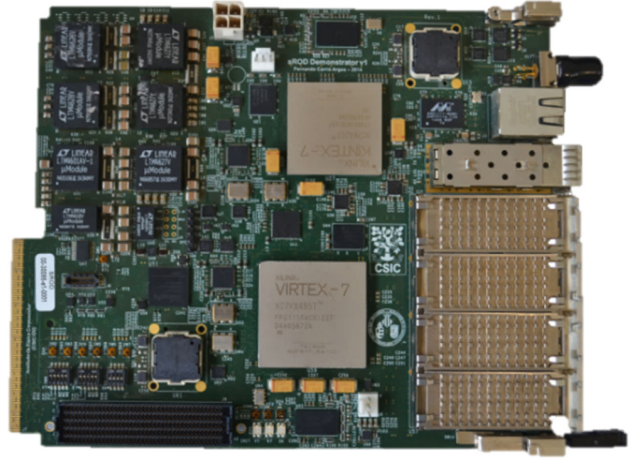

Fig. 6. Picture of the TilePPr Demonstrator.

## IV. THE TILEPPR DEMONSTRATOR

The TilePPr Demonstrator board [8] was designed to acquire and process the detector data transmitted from the Demonstrator module, as well as for the distribution of the LHC clock and Timing, Trigger and Commands (TTC) to the on-detector electronics. This prototype serves one TileCal module representing $1/8^{th}$ of the final TilePPr module for the HL-LHC.

The TilePPr Demonstrator board is a double mid-size Advanced Mezzanine Card (AMC), which can be operated in an Advanced Telecommunications Computing Architecture (ATCA) carrier or in a Micro Telecommunications Computing Architecture (µTCA) crate. The PCB counts 16 layers, where 8 layers are devoted for power distribution and ground planes, and 8 layers are used for data lines. NELCO 4000-13 SI dielectric material was selected to reduce high frequency losses in the high-speed lines interfacing with the on-detector electronics.

This prototype is equipped with four QSFP modules connected to one Virtex 7 XC7VX485T FPGA (Readout FPGA), one Kintex 7 XC7K420T FPGA (Trigger FPGA), two Texas Instruments CDCE62005 jitter cleaners. It also hosts DDR3 memories and one Ethernet port in the front panel. The AMC backplane provides power to the TilePPr, as well as four point-to-point high-speed connections between the Readout FPGA and the Rear Transition Module (RTM), and several GbE and PCIe ports to connect both FPGAs with rest of the boards in the ATCA shelf.

The Readout FPGA implements all the firmware required for the readout and operation of the Demonstrator module, and the Trigger FPGA preprocesses and transmits data to the upgraded trigger system. Figure 6 shows a picture of the TilePPr Demonstrator board.

### A. TilePPr validation tests

As part of the validation process of the TilePPr prototype, extensive Bit Error Rate (BER) tests and jitter measurements were performed to evaluate the signal quality of the clock and data transmitted towards the on-detector electronics.

These jitter measurements are crucial for the validation of this high-speed readout system, since the jitter affects the performance of all the components involved in the data communication chain, such as clocks, jitter cleaners, high-speed transceivers and power supplies.

Different types of jitter data were extracted from the optical output of the QSFP modules using a Keysight DCA-X 86100D sampling oscilloscope [9] equipped with an optical module 86105C. The optical signal quality was measured at the output of an Avago AFBR-79Q4Z QSFP module by connecting to the oscilloscope a multimode fiber one meter long. The transceivers were configured to operate the GBT links at 4.8 Gbps and 9.6 Gbps without any adjustment of the pre- or post-emphasis features.

Table II presents the jitter measurement results (µ) done at 4.8 Gbps ($T_{bit}\approx 208$ ps) and 9.6 Gbps ($T_{bit}\approx 104$ ps) data rates with the corresponding standard deviation (σ) of the measurement. The obtained jitter measurements show a good signal quality with low jitter values for the communication towards the on-detector electronics.

TABLE II
AVERAGE (µ) AND RMS (σ) VALUES OF THE RJ, DJ, AND TJ FOR DIFFERENT TRANSMISSION RATES AND DIFFERENT BER VALUES

|  | 4.8 Gbps | | 9.6 Gbps | |
|---|---|---|---|---|
|  | µ (ps) | σ (ps) | µ (ps) | σ (ps) |
| RJ (RMS) | 2.74 | 0.28 | 2.97 | 0.25 |
| DJ(δ- δ) | 2.44 | 1.31 | 5.94 | 1.29 |
| TJ($10^{-12}$) | 39.85 | 3.46 | 46.50 | 3.15 |
| TJ($10^{-14}$) | 43.31 | 3.16 | 50.25 | 2.88 |
| TJ($10^{-16}$) | 46.51 | 2.83 | 53.71 | 2.59 |
| TJ($10^{-18}$) | 49.50 | 2.48 | 56.95 | 2.27 |

The estimated Total Jitter (TJ) values for the given BER indicate that the design is robust enough to operate at the required data rates. The estimated TJ ($10^{-18}$) corresponds to 0.23 Unit Interval (UI) for 4.8 Gbps and to 0.55 UI for 9.6 Gbps, both at a BER of $10^{-18}$. The Random Jitter (RJ) presents similar values for 4.8 Gbps and 9.6 Gbps data rates, while the Deterministic Jitter (DJ) indicates a higher jitter at 9.6 Gbps. The Keysight DCA-X 86100D provided a resolution of 5 fs and 2 fs for the RJ measurement at 4.8 Gbps and 9.6 Gbps respectively, and 50 fs and 20 fs for the rest of the measurements. Figure 7 shows the eye diagrams obtained with the sampling scope during the test, resulting in wide open eyes in time and amplitude at both data rates.

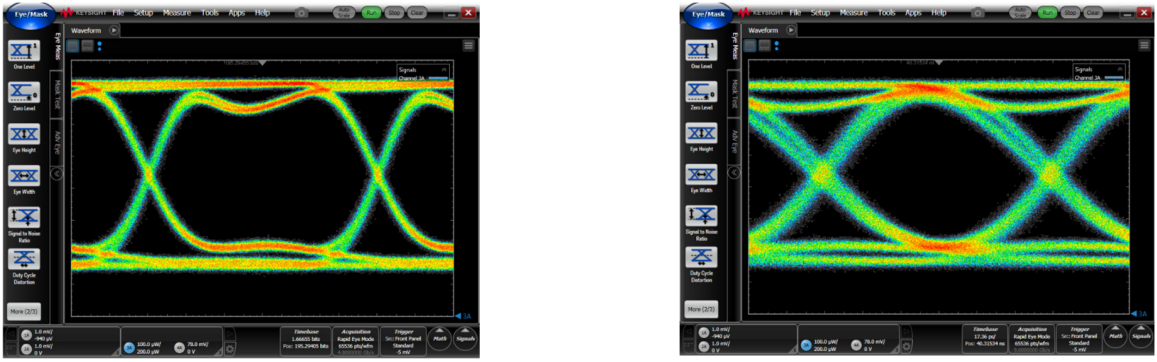

Fig. 7. Optical eye diagrams measured with a Keysight DCA-X 86100D oscilloscope at 4.8 Gbps (left) and at 9.6 Gbps (right).

Finally, BER tests were performed using the IBERT IP core [10] on the sixteen links running at 9.6 Gbps with a PRBS-31 data pattern over a period of 115 hours. No errors were found during the tests corresponding to a BER $\leq 5\cdot 10^{-17}$ with a confidence level of 95%.

## V. Clock and readout architecture at the test beam

The clock and readout architecture implemented for the readout and operation of the Demonstrator module is very close to the final one proposed for the HL-LHC. The difference between the Demonstrator and HL-LHC readout architectures lies in the capability of Demonstrator module to provide analog trigger signals to the ATLAS Level-1 trigger system. This feature is required to operate the Demonstrator module within the current DAQ architecture of ATLAS, if the Demonstrator module will be inserted in the TileCal detector to take data during the run 3 (2021-2023). Figure 8 shows a complete block diagram with all the components and interconnections of the readout architecture employed during the test beam campaigns.

### A. Readout architecture at the test beam setup

The bi-directional communication between the TilePPr and the Demonstrator module is implemented using the GigaBit Transceiver (GBT) protocol [11] with an asymmetric bandwidth (4.8 Gbps / 9.6 Gbps). The required sixteen GBT links transmit Detector Control System (DCS) and TTC commands to the on-detector electronics at 4.8 Gbps, and receive detector data in the TilePPr at 9.6 Gbps through the four QSFP modules. The sampling clock is transmitted to the on-detector electronics embedded with the TTC and DCS data through the GBT links.

In the on-detector electronics, each Daughter Board receives four links from the TilePPr corresponding to one QSFP module. One of the receiver links is connected to the input of a GBTx chip [12] for the recovery of the reference clock and for the remote configuration of the on-detector electronics. The remaining three GBT links are connected to two Kintex 7 FPGAs for the reception of configuration and control commands.

The Daughter Board is plugged into a Main Board through a 400-pin FMC connector. This connector provides a high-speed path to receive the PMT digitized data and permits the configuration of the Main Board components. Both Daughter Board and Main Board are divided in two independent halves called A and B-sides, which are individually powered with 10 V from the Low Power Voltage Supplies (LVPS) for redundancy.

The Main Board is divided in four sections each controlled by an Altera Cyclone IV FPGA. A section contains the required circuitry to control and read out 3 PMTs for a total of 12 PMTs per Main Board. Each Cyclone IV controls 3 dual-channel 12-bit ADCs for digitizing the shaped PMT signals provided by the 3-in-1 cards (high- and low-gain) at 40 Msps, 6 DACs to control the bias voltage levels applied to the ADC inputs and 3 slow-speed ADCs for sampling the integrators at 50 ksps. The Cyclone FPGAs are accessed from the Daughter Board via an SPI interface, while digitized PMT signals are sent directly from the ADCs to the Daughter Board using LVDS lines at 560 Mbps. Two I2C buses (one per side) are dedicated for the readout of the integrator ADCs.

In the Daughter Board, the serial data coming from the 12 high-speed ADCs at 560 Mbps are deserialized with fixed and deterministic latency. The ADC data are packed with the integrator and monitoring data and transmitted to the TilePPr Demonstrator board through a pair of redundant GBT links per side running at 9.6 Gbps.

The TilePPr receives the digitized data from four Daughter Boards at 40 MHz through the four QSFP modules requiring a total input bandwidth of 160 Gbps. Data are unpacked and stored in the circular pipeline memories of the Readout FPGA implemented with dedicated block RAM resources upon the reception of a Level-1 trigger Acceptance (L1A) signal. When the L1A signal is received, the TilePPr collects the data for the selected event from the pipelines and transmits them to the FELIX system (16 samples and 2 gains) and to the legacy ROD (7 samples and 1 gain) with the proper format, keeping backward compatibility with the current ATLAS DAQ system.

### B. Clock distribution at the test beam setup

Related to the distribution of the sampling clock towards the on-detector electronics, the TilePPr recovers the 40 MHz clock from the TTC stream with an Analog Devices ADN2814 chip [13]. The recovered clock is buffered to a CDCE62005 jitter cleaner to meet the jitter requirements of the transceiver reference clock [14]. As explained above, the 40 MHz clock is recovered in the Daughter Boards and distributed as sampling clock to the Main Board, where the Cyclone IV FPGAs fan out the clock to the ADCs.

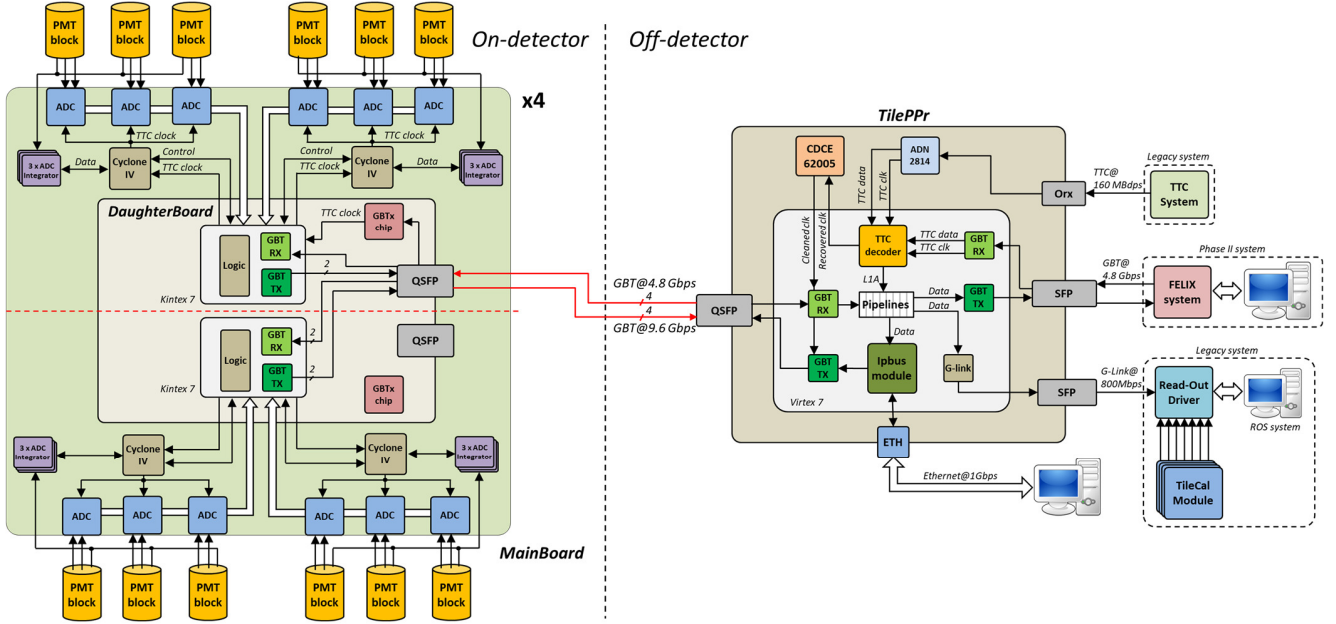

Fig 8. Block diagram of the readout architecture of the Demonstrator module during the test beam campaigns.

An important feature of this system is that the GBT links were configured for fixed and deterministic latency operation, minimizing the phase variations in the distributed clock after power cycling the readout electronics.

This operation mode of the GBT links permits the correct assignment of the bunch crossing identifier to the digitized data in the TilePPr, since the digitized data is not time stamped in the on-detector electronics. For this reason, the transceivers of the Daughter Board FPGAs use the recovered clock from the GBTx chip for the transmission to the off-detector electronics, operating both on- and off-detector electronics systems in the same clock domain.

All the data flow and control of the TilePPr and configuration commands for the on-detector electronics is implemented over an Ethernet network using the IPbus protocol [15]. A set of IPbus registers are used to handle the TTC and DCS commands to the on-detector electronics as well as for data monitoring and calibration. The internal configuration for the different modes of operation and remote monitoring of the TilePPr status is managed through the IPbus registers.

## VI. CLOCK STABILITY TESTS

The stability of the clock distributed towards the on-detector electronics was studied using digital phase monitoring tools implemented in the logic of the Readout FPGA. The quality of the clock signal distributed as a sampling clock plays a crucial role in the energy and time reconstruction of the physics events produced in the detector.

The phase monitoring module implemented in the TilePPr permits the detection and tracking of non-deterministic latency variations in the high-speed path between on- and off-detector electronics. The jitter cleaners or FPGA transceivers could introduce a latency variation after power cycle, or due to temperature and voltage drifts.

### A. OverSampling to UnderSampling circuit

Each one of the sixteen GBT links was connected to a phase monitoring module composed of one OverSampling to UnderSampling circuit (OSUS) [16]. The OSUS circuit tracks the phase difference between the 40 MHz clock transmitted and recovered back from the on-detector electronics.

The OSUS circuit is a digital circuit based on the Digital Dual Mixer Time Domain (DDMTD) circuit [17] [18], but it combines both subsampling and oversampling techniques. The combination of both sampling techniques makes possible the measurement of clock phase differences between signals with frequencies different from the one used to generate the sampling clock. In addition, this technique permits to increase the acquisition rate of phase measurements with respect to the original DDMTD circuit.

Figure 9 presents a functional block diagram of the OSUS circuit. The sampling clock ($u_s$) is generated from the first input clock ($u_1$) using a Phase Locked Loop (PLL). The PLL distributes to the samplers a sampling clock $u_s$ with a frequency of $M \cdot \frac{N}{N+1}$ times the input clock frequency, where M is the oversampling factor and N defines the resolution of the OSUS circuit. M and N factors are natural numbers.

Both $u_1$ and $u_2$ input clocks are sampled with $u_s$, obtaining the oversampled $U_1$ and $U_2$ signals. These oversampled signals are passed to the Phase Control block, which multiplexes the oversampled signals in intervals of M·$T_s$ and builds M pairs of subsampled $U_n^k$ signals.

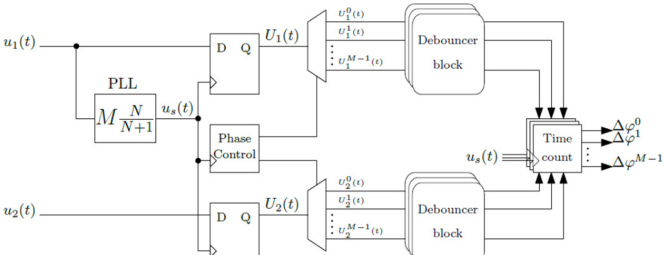

Fig. 9. Block diagram of the OSUS circuit.

Figure 10 shows the timing diagram of an OSUS circuit with M=4 and N=8, representing the sampling clock ($u_s$), one input clock ($u_1$), its corresponding oversampled signal ($U_1$), and posterior regrouped samples forming the subsampled signals ($U_1^k$).

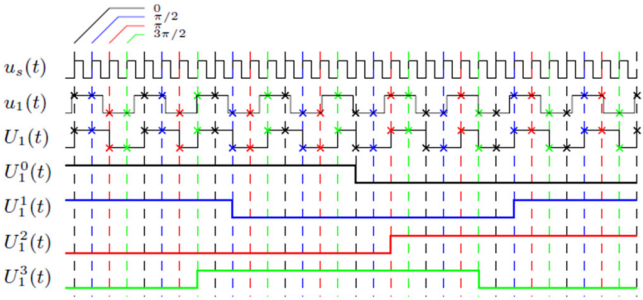

Fig. 10. Timing diagram of the OSUS circuit signals for M=4 and N=8.

The Debouncer block filters any possible glitch in the subsampling signals before passing them to the time counters, and it estimates the edge position of the $U_1^k$ and $U_2^k$ signals. The phase difference $\Delta\varphi^k$ between both input clocks is calculated as indicated in (1).

$$\Delta\varphi^k = \frac{1}{N+1} \cdot T_s \cdot n_{cycles}^k \quad (1)$$

where $n_{cycles}^k$ corresponds to the number of clock cycles of $u_s$ between the rising edges of the filtered $U_n^k$ signals with the same index $k$.

B. *Implementation in the Readout FPGA*

All phase monitoring modules implemented in the Readout FPGA drive the same sampling clock $u_s$ to the samplers. Thus, a single Clock Management Tile (CMT) [19] block is required to generate a 240 MHz sampling clock $u_s$ from the 40 MHz clock recovered from the TTC system.

The chosen N factor is 16384, leading into a resolution of ~1.5 ps for the measurement of 40 MHz clock signals (M = 6). However, the accuracy of the phase monitoring module was measured to be ~30 ps$_{RMS}$ during the tests. The accuracy is limited by the jitter of the clock extracted from the transceiver, and by the jitter introduced by the signal routing through the fabric logic [20] and clock buffers [19].

Figure 11 and 12 present the results of the studies on the stability of the clock distributed clock from the TilePPr towards the on-detector electronics, where $u_1$ signal corresponds to the clock used for transmission and $u_2$ signal to the clock recovered from the on-detector electronics.

During the tests the phase difference between clocks was acquired after resetting the Daughter Board 100 times. After each reset, 1,000 phase measurements between $U_1^0$ and $U_2^0$ were taken and read out through the IPbus interface. Then, the average value was calculated in the computer for the set of measurements.

Figure 11 shows the histogram of all phase measurements taken during the tests for the four channels (A0, A1, B0 and B1) of one QSFP module. The phase difference depends on the propagation delay introduced by the fiber optic, optoelectronic modules, FPGA resources and routing. As can be observed in Figure 11, the results do not show significant variations between the phase of the distributed and recovered clock. Thus, the operation of the GBT links with fixed and deterministic latency is validated.

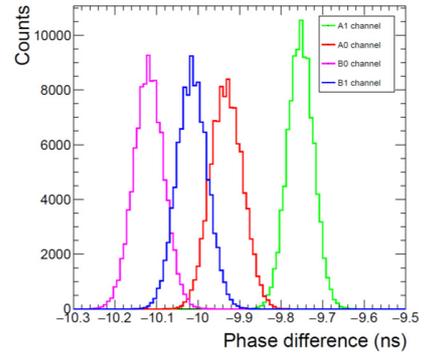

Fig. 11. Histogram of the phase measurements taken after resetting the on-detector electronics 100 times.

Figure 12 presents the average value of 1,000 measurements after each reset as a function of the reset count. As shown in Figure 12, the maximum time variation between both clocks after resetting the on-detector electronics is 100 ps$_{pk-pk}$. Such variations in the phase of the sampling clock have no impact in the detector performance in terms of energy and time reconstruction resolution [22], as it was observed during the data-taking period in Run 1 (2010-2012). Hence, the proposed clock distribution strategy fulfills the requirements for the HL-LHC.

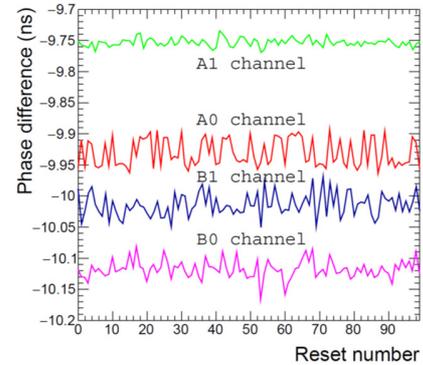

Fig. 12. Averaged value of the 1,000 phase measurements taken after each reset as a function of the reset count.

## VII. CONCLUSIONS

The ATLAS Tile Calorimeter plans a complete redesign of the readout electronics for the HL-LHC using a new readout strategy with a full-digital trigger. In the new readout architecture, the on-detector electronics will transmit all the digitized samples to the off-detector electronics at the LHC frequency. The sampling clock for the digitization of the PMT signals will be distributed embedded with the data from the off-detector electronics.

A Demonstrator module was built with prototypes of the upgraded electronics, using the new readout and clock distribution strategy for the HL-LHC. The Demonstrator module was tested in several test campaigns with particle beams between 2015 and 2018, where the on-detector electronics transmitted the digitized detector signals at 40 MHz to the TilePPr module.

Different tests were performed to validate the proposed readout and clock distribution strategy for the TileCal detector at the HL-LHC. Digital phase monitoring circuits implemented in the TilePPr were used to detect small phase deviations in the distributed clock produced after resetting the on-detector electronics. The stability of the clock ensures a good performance of the algorithms for energy and time reconstruction in TileCal.

## ACKNOWLEDGMENTS

Authors would like to thanks all the TileCal Collaboration members who have contributed to the development of the TileCal Demonstrator module.

## REFERENCES

[1] ATLAS Collaboration, "Readiness of the ATLAS Tile Calorimeter for LHC collisions," in *European Physics Journal C*, Vol. 70, Issue 4, pp. 1193-1236, December 2010.
[2] ATLAS Collaboration, "The ATLAS Experiment at the CERN Large Hadron Collider," 2008 *JINST* 3 S08003.
[3] A. Valero et al., "ATLAS TileCal ReadOut Driver production," 2007 *JINST* 2 P05003.
[4] ATLAS Collaboration, "Technical Design Report for the Phase-II Upgrade of the ATLAS Tile Calorimeter," CERN-LHCC-2017-019, 2017.
[5] J. Anderson et al., "FELIX: a High-Throughput Network Approach for Interfacing to Front End Electronics for ATLAS Upgrades," in *Journal of Physics: Conference Series* 664 (2015) No.8, 082050.
[6] F. Tang et al., "Design of the front-end readout electronics for the ATLAS tile calorimeter at the sLHC," in *IEEE Transactions on Nuclear Science*, Vol. 60, No. 2, pp. 1255-1259, April 2013.
[7] S. Muschter et al., "Development of a digital readout board for the ATLAS Tile Calorimeter upgrade demonstrator," 2014 *JINST* 9 C01001.
[8] F. Carrió, P. Moreno and A. Valero, "Performance of the Tile PreProcessor Demonstrator for the ATLAS Tile Calorimeter Phase II Upgrade," 2016 *JINST* 11 C03047.
[9] Keysight Technologies, "Infiniium DCA-X 86100D Wide-Bandwidth Oscilloscope Mainframe and Modules," DCA-X 86100D datasheet, August 2016.
[10] Xilinx Inc., "Integrated Bit Error Ratio Tester 7 Series GTX Transceivers v3.0," PG132, June 2016.
[11] M. Barros et al., "The GBT-FPGA core: features and challenges," 2015 *JINST* 10 C03021.
[12] C. Ghabrous Larrea et al., "IPbus: a flexible Ethernet-based control system for xTCA hardware," 2015 *JINST* 10 C02019.
[13] Analog Devices, "Continuous Rate 10 Mb/s to 675 Mb/s Clock and Data Recovery IC with Integrated Limiting Amp," ADN2814 datasheet rev. C, 2012.
[14] Silicon Labs, "FPGA Reference Clock Phase Jitter Specifications," Application Note AN699 rev. 0.1, 2012.
[15] P. Moreira, J. Christiansen and K. Wyllie, "GBTx Manual – GBT project," Manual v0.15, 2016.
[16] F. Carrió, "Development of Readout Electronics for the ATLAS Tile Calorimeter at the HL-LHC," PhD Thesis, Valencia, May 2017.
[17] P. Moreira, P. Alvarez, J. Serrano, I. Darwezeh and T. Wlostowski, "Digital dual mixer time difference for sub-nanosecond time synchronization in Ethernet," in *2010 IEEE International Frequency Control Symposium*, pages 449–453, June 2010.
[18] B. Amrutur, P.K. Das and R. Vasudevamurphy, "0.84 ps Resolution Clock Skew Measurement via Subsampling," in *IEEE Transactions on Very Large Scale Integration (VLSI) Systems*, Vol. 19, No. 12, pp. 2267-2275, December 2011.
[19] Xilinx Inc., "7 Series FPGAs Clocking Resources v1.13," UG472, March 2017.
[20] A. Aloisio, F. Cevenini, R. Giordano and V. Izzo, "Characterizing Jitter Performance of Multi Gigabit FPGA-Embedded Serial Transceivers," in *IEEE Transactions on Nuclear Science*, Vol. 57, No. 2, pp. 451-455, April 2010.
[21] A. Valero, "Implementation and Performance of the Signal Reconstruction in the ATLAS Hadronic Tile Calorimeter," in *Physics Procedia*, Vol. 37, pp. 1765-1771, 2012.
[22] The ATLAS Collaboration, "Operation and performance of the ATLAS Tile Calorimeter in Run 1," Submitted to the *European Physical Journal C* in June 2018.